\documentclass[aps,prx,twocolumn,superscriptaddress,showpacs,floatfix]{revtex4-2}
\usepackage{color}
\usepackage{mathtools}
\usepackage[c]{esvect}
\usepackage{bm}        
\usepackage{amssymb}   
\usepackage{amssymb, amsthm, float, graphicx, amsfonts, dsfont}
\usepackage{amsmath} 

\usepackage[colorlinks=true,citecolor=blue,linkcolor=magenta, breaklinks=true]{hyperref}
\usepackage{graphicx}
\usepackage{braket}
\usepackage{natbib}
\usepackage [latin1]{inputenc}
\usepackage{siunitx}
\usepackage{blkarray, bigstrut}
\usepackage[dvipsnames]{xcolor}
\usepackage{comment}
\usepackage{physics}
\usepackage{tikz}

\newcolumntype{P}[1]{>{\centering\arraybackslash}p{#1}}
\newcommand{\bea}{\begin{eqnarray}}
\newcommand{\eea}{\end{eqnarray}}

\newcommand{\br}{\mathbf{r}}

\newcommand{\be}{\begin{equation}}
\newcommand{\ee}{\end{equation}}

\newcommand{\beal}{\begin{align}}
\newcommand{\eeal}{\end{align}}
\newcommand{\ra}{\rangle}
\newcommand{\la}{\langle}

\renewcommand{\vec}[1]{\mathbf{#1}}

\newcommand{\btjstrw}{\mathrel{{\rotatebox[origin=c]{90}
{$\bowtie$}}\kern-0.18em\raisebox{-.95ex}{$\bullet$}
\kern-0.5em\raisebox{.97ex}{$\bullet$}
\kern-1.12em\raisebox{.97ex}{$\bullet$}
\kern-0.52em\raisebox{-.95ex}{$\bullet$}}}

\newcommand{\btjnbrR}{{\mathrel{\rotatebox[origin=c]{90}
{$\bowtie$}}\kern-0.22em\raisebox{.9ex}{$\bullet$}
\kern-1.em\raisebox{-.8ex}{$\bullet$}}}
\newcommand{\btjnbrL}{{\mathrel{\rotatebox[origin=c]{90}
{$\bowtie$}}\kern-0.22em\raisebox{-.8ex}{$\bullet$}
\kern-1.em\raisebox{+.9ex}{$\bullet$}}}

\def\bco{Ba$_2$CaOsO$_6$}
\def\bnco{Ba$_2$Ca$_{1-\delta}$Na$_\delta$OsO$_6$}
\def\bsco{Ba$_2$Ca$_{1-\delta}$Sr$_\delta$OsO$_6$}
\def\bcmo{Ba$_2$Ca$_{1-\delta}$Mg$_\delta$OsO$_6$}

\begin{document}

\title{Probing octupolar hidden order via Janus impurities}
\author{Sreekar Voleti}
\affiliation{Department of Physics, University of Toronto, 60 St. George Street, Toronto, ON, M5S 1A7 Canada}
\author{Koushik Pradhan}
\affiliation{Department of Condensed Matter Physics and Materials Science, S.N. Bose National Centre for Basic Sciences, Kolkata 700098, India.}
\author{Subhro Bhattacharjee}
\affiliation{International Centre for Theoretical Sciences, Bengaluru 560089, India}
\author{Tanusri Saha-Dasgupta}
\affiliation{Department of Condensed Matter Physics and Materials Science, S.N. Bose National Centre for Basic Sciences, Kolkata 700098, India.}
\author{Arun Paramekanti$^*$}
\affiliation{Department of Physics, University of Toronto, 60 St. George Street, Toronto, ON, M5S 1A7 Canada}
\affiliation{Department of Condensed Matter Physics and Materials Science, S.N. Bose National Centre for Basic Sciences, Kolkata 700098, India.}
\affiliation{International Centre for Theoretical Sciences, Bengaluru 560089, India}
\email{arun.paramekanti@utoronto.ca}
\date{\today}
\begin{abstract}
{\bf {Quantum materials with non-Kramers doublets are a fascinating venue to realize multipolar 
hidden orders.
Impurity probes which break point group symmetries, such as implanted muons or substitutional impurities,
split the non-Kramers degeneracy and exhibit a Janus-faced influence in such systems: they can destroy the very
order they seek to probe. Here, we explore this duality in
cubic osmate double perovskites which are candidates for exotic $d$-orbital octupolar order competing
with quadrupolar states.
Using {\it ab initio} computations, Landau theory, and Monte Carlo simulations, we show that Janus
impurities induce local strain fields, nucleating quadrupolar puddles and suppressing the octupolar $T_c$.
At the same time, strains mix the non-Kramers doublet with an excited magnetic
triplet, creating parasitic dipole moments which directly expose the hidden octupolar order parameter.
Our work unravels this Janus duality in recent
impurity nuclear magnetic resonance (NMR) experiments, with important implications for 
uncovering hidden order in diverse multipolar materials.}}
\end{abstract}

\maketitle

The idea of ``hidden order'', or subtle broken symmetries which are difficult to detect, has a long history in 
$f$-electron heavy fermion materials, with notable examples such as NpO$_2$ and URu$_2$Si$_2$
\cite{SantiniNpO2_PRL2000,Tripathi2002,NpO2TripleQ_PRL2002,Fazekas_PRB2003, NpO2NMR_PRL2006,Coleman2007,MultipolarRMP2009,HauleKotliar2009,Sakai_JPSJ2011,Sato_PRB2012,Rau2012,Flint2013,Arima2013,
Nakatsuji_PRL2014,Blumberg_Science2015,Hattori2016,Freyer2018,SBLee2018,Patri2019, Jang_2017_npjqm}.
The intricacy of these orders stems
from their spin-orbital entanglement and complex multipolar character.
Understanding the origin of such multipolar orders and devising probes to characterize them
remain important challenges in this field.
Recent work has discovered multipolar orders in metallic and insulating
$d$-orbital systems
\cite{ChenBalents2010,ChenBalents2011,LFu_PRL2015,Hsieh_Science2017,Motome2018,svoboda2021,Mitrovic_NComm2017,Mitrovic_Physica2018,Hiroi_JPSJ2019,svoboda2021,maharaj2019octupolar,paramekanti2019octupolar,voleti2020,Lovesey2020,pourovskii2021,Kee2021,Voleti2021,
Churchill2022},
opening up the exploration of distinct classes of materials and probes to shed light on these complex orders.

The simplest nontrivial multipolar order involves ordering of quadrupoles, and it is intimately tied to the
breaking of crystalline rotational symmetry, i.e. `nematic order'.
Such nematic phases and nematic quantum critical points 
are of great interest in cuprate, iron pnictide,
and iron chalcogenide materials \cite{NematicReview,NematicReviewFernandes}. 
Experiments on heavy Mott insulators with strong spin-orbit coupling (SOC)
have also revealed candidates hosting quadrupolar orders.
For Mott insulators with one electron in the $t_{2g}$ orbital, SOC
leads to a $j\!=\! 3/2$ pseudospin which supports higher multipole operators 
\cite{ChenBalents2010,ChenBalents2011,svoboda2021}.
High resolution X-ray diffraction experiments on a candidate material, Ba$_2$MgReO$_6$, have found
compelling evidence for separate phase transitions associated with ordering of the associated
quadrupole and dipole moments on Re
\cite{Hiroi_JPSJ2019,Hiroi_PRR2020}. Earlier NMR experiments had proposed such successive broken
symmetries in Ba$_2$NaOsO$_6$ \cite{Mitrovic_NComm2017,Mitrovic_Physica2018}. Glassy valence-bond states,
possibly with short-range multipolar orders, may be relevant to $j\!=\! 3/2$ Ba$_2$LuMoO$_6$ \cite{d1_Mustonen_npjQM2022}.

\begin{figure}[t]
\centering
    \includegraphics[width=0.45\textwidth]{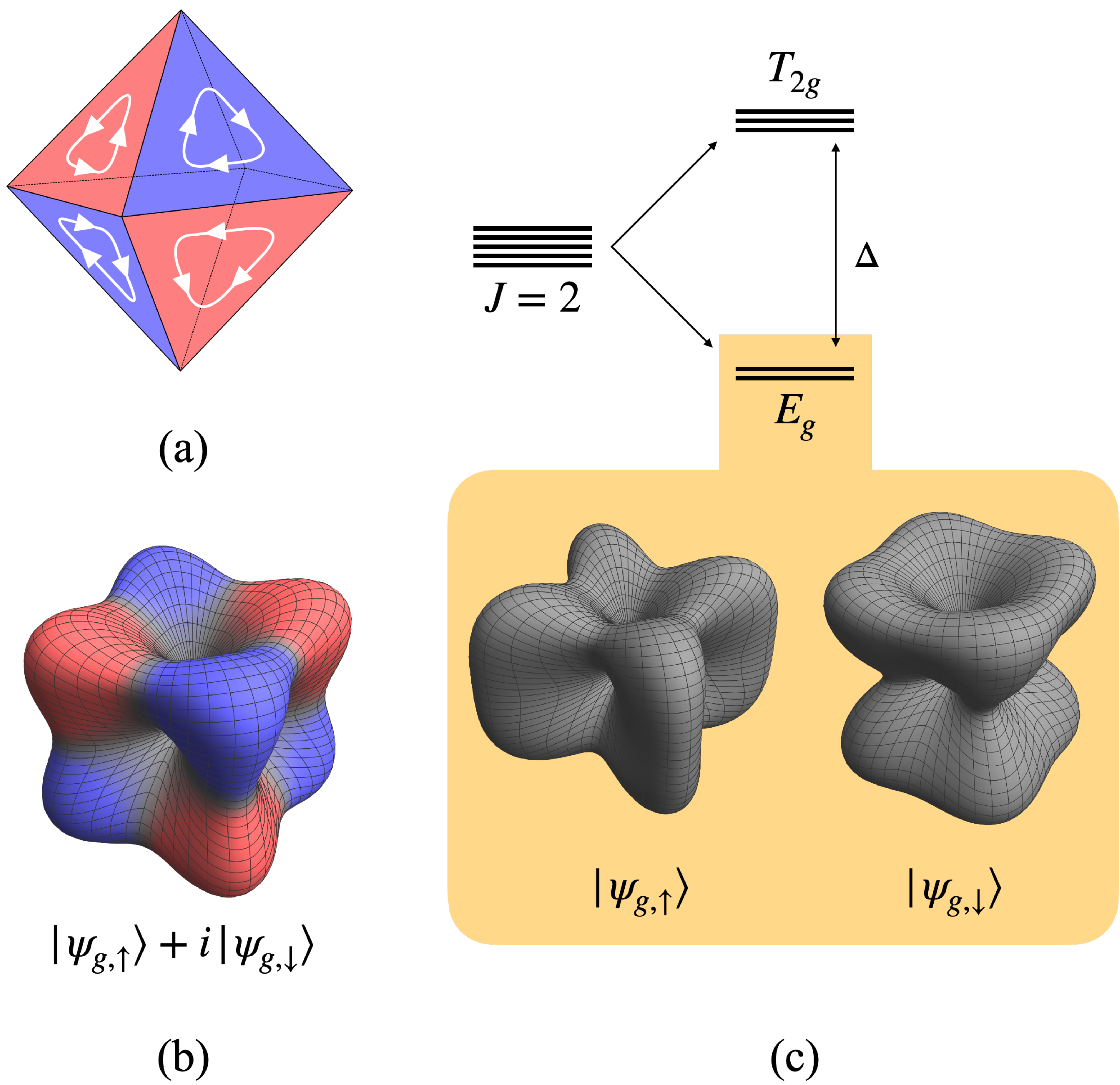}
   \caption{ {\bf Ising octupolar order.} {\bf a.} Schematic view of octupolar order as arising from an
    alternating pattern of orbital currents around transition metal octahedra. {\bf b.} Octupolar order for $t_{2g}$ orbitals, with shape corresponding to the electron charge density cloud (which preserves
    cubic symmetry) and colors denoting the magnetization density due to orbital currents. {\bf c.} Microscopic origin of multipole orders in $J\!=\! 2$ magnets with SOC. 
    The octahedral crystal field weakly splits the $J\!=\!2$ levels into a low energy many-body 
    $E_g$ non-Kramers doublet and an excited $T_{2g}$ magnetic triplet. 
    We highlight the charge density in the real eigenfunctions (``orbitals'')
    $|\psi_{g,\uparrow}\rangle$ and $|\psi_{g,\downarrow}\rangle$ 
    of the $E_g$ levels; see text for details. The octupolar order in {\bf b} 
    is a complex superposition of these real wavefunctions. The time-reversed partner $|\psi_{g,\uparrow}\rangle - i |\psi_{g,\downarrow}\rangle$ (not shown) has reversed loop currents and magnetization densities.}
        \label{fig:oct_schematic}
\end{figure}

Going one step further, octupolar ordering corresponds to a pattern of loop current
order which breaks time-reversal symmetry and leads to an octupolar magnetization density
as depicted in Figs.~\ref{fig:oct_schematic}{\bf a} and \ref{fig:oct_schematic}{\bf b}.
{Ferro-octupolar order is a spin-orbit coupled variant of `altermagnetism' 
\cite{Mazin2022,Sinova2022,Spaldin2022}, which has received a lot of recent attention.}
Closely related loop currents have been discussed in cuprates \cite{Hsu1991,DDWcuprates,VarmaLoop}, 
manganites \cite{khomskii2001}, iridates \cite{Hsieh2016,Bourges_2022_IridateLoopCurrents}, and the 
recent kagome superconductors \cite{Morten2022_KagomeLoopCurrents,Mielke_2022_KagomeLoopCurrents}; however, 
smoking-gun signatures of these currents remain to be observed. Chiral-lattice antiferromagnets
such as Pb(TiO)Cu$_4$(PO$_4$)$_4$ may also host octupoles \cite{chiral_octupole_Kimura_npjQM2021}.

Quantum materials which host non-Kramers doublets 
provide
the most promising venue to search for multipolar orders \cite{Onoda2010,SBLee2012,hattori2014antiferro,Kadowaki2016,Patri_PRR2020,onimaru2016exotic,Seth2022ProbingEQ}. 
Unlike Kramers doublets, the degeneracy of non-Kramers doublets is protected by crystalline point
group symmetries. These pseudospin-$1/2$ degrees of freedom thus cannot reflect pure dipoles
since their degeneracy can be lifted by breaking crystal symmetries while preserving time-reversal symmetry.
The ordering of these non-Kramers doublets, facilitated by exchange interactions, results 
in multipolar symmetry breaking. 

{An important route to sleuthing broken symmetries in quantum materials 
is via local probes such as NMR of doped impurity ions or muon spin rotation ($\mu$SR) of implanted muons.
In the context of multipolar orders, we coin the term `Janus impurities' to refer
to such probes since they are two-faced 
like the Roman God {\it Janus}: while they may act as local probes of multipolar order,
these impurities also break the local point group symmetry and split the non-Kramers degeneracy, 
thus destroying the very order they seek to detect. In this work, we ask whether 
Janus impurities can nevertheless be useful probes of multipolar orders.}

As an example, recent work has shown that non-Kramers doublets naturally appear in heavy $d$-orbital Mott insulators, 
with strong SOC, which host two electrons in the $t_{2g}$ orbital
\cite{maharaj2019octupolar,paramekanti2019octupolar,voleti2020,Voleti2021}. 
In these systems, the $j$-$j$ coupling between the two $j\!=\!3/2$ electrons leads to a $J\!=\!2$ multiplet, 
which splits in the cubic environment to yield
a low energy non-Kramers doublet, as shown in Fig.~\ref{fig:oct_schematic}{\bf c}; 
we denote these doublet wavefunctions by $|\psi_{g,\uparrow}\rangle$ and $|\psi_{g,\downarrow}\rangle$ 
For these doublets, the pseudospin-$1/2$ operators
$(\tau_x,\tau_z)$ transform as a two-dimensional electric quadrupole, while $\tau_y$ transforms as a magnetic 
octupole \cite{maharaj2019octupolar,paramekanti2019octupolar,voleti2020}. 
Ordering of $(\tau_x,\tau_z)$ and of $\tau_y$ thus correspond respectively
to quadrupolar and Ising-type octupolar orders.
Ferro-octupolar order with $\la \tau_y \ra \!\neq \! 0$
naturally explains \cite{maharaj2019octupolar,paramekanti2019octupolar,voleti2020,Voleti2021}  
the time-reversal breaking transition seen via $\mu$SR
in the osmium
double perovskite (DP) Mott insulators, Ba$_2$CaOsO$_6$, Ba$_2$MgOsO$_6$, and Ba$_2$ZnOsO$_6$
\cite{Thompson_JPCM2014,Kermarrec2015,Thompson_PRB2016,MarjerrisonPRB2016,maharaj2019octupolar}
where Os$^{6+}$ ions host a $5d^2$ configuration.

Microscopic support for ferro-octupolar order in the cubic osmate DPs comes from recent
density functional
and dynamical mean field theory (DFT+DMFT) calculations \cite{pourovskii2021}, as well as
model tight-binding calculations which extract the intersite exchange in
the presence of electron interactions and SOC
\cite{Voleti2021,Churchill2022}. Raman scattering from crystal
field levels is one proposed route to probe such octupolar order \cite{paramekanti2019octupolar}; however, 
smoking gun experimental signatures of this exotic order are still lacking. Furthermore,
while octupolar order is the most likely broken symmetry state in these DPs,
there can be
competing quadrupolar orders \cite{Kee2021,Voleti2021,Churchill2022}.
Such competing orders are important at crystal surfaces
where the cubic symmetry is naturally
broken by uniaxial strain fields, favoring quadrupolar order \cite{Voleti2021}.

While `hidden order' has been discussed in the literature, most efforts to date are 
theoretical studies of toy models, and clear evidence of hidden octupolar order has so far remained elusive.
Here, we use
{\it ab initio} calculations of the crystal structure, Landau theory analysis,
Monte Carlo (MC) simulations, and modelling of NMR spectra, to
explore the competition 
between octupolar and quadrupolar orders in the osmate double perovskites.
Our work uncovers the evidence for octupolar order by making direct contact with recent experimental studies.

For Ba$_2$CaOsO$_6$, our {\it ab initio} computation of the phonon modes shows 
that the cubic $Fm\bar{3}m$ 
structure is stable against deformations. This bolsters the case for the possibility of
octupolar ordering which preserves
cubic symmetry, since non-cubic lattice distortions are otherwise likely to favor competing
quadrupolar orders \cite{Kee2021}. We find a low energy peak in the phonon density of states (DOS) 
which is in good agreement with inelastic neutron scattering experiments \cite{maharaj2019octupolar}.
Upon substituting Ca$^{2+}$ with Na$^+$ or Sr$^{2+}$, we find that these Janus impurities
lead to local distortions in the crystal which
break the octahedral point group symmetry around Os. Such strain fields acts as local transverse fields on the
octupolar order, creating droplets of pinned 
quadrupolar order. 

Our Monte Carlo simulations show that these 
quadrupolar droplets strongly suppress the ferro-octupolar $T_c$, and
lead to an inhomogeneous and temperature-dependent distribution of octupolar and quadrupolar moments.
Furthermore, we show that these strain fields can reveal the intrinsic octupolar order through the 
generation of parasitic dipole moments below the ferro-octupolar $T_c$.
Finally, by explicitly modelling the $^{23}$Na NMR spectrum, we show how our work
allows us to capture
all the key features observed in recent $^{23}$Na NMR experiments on {\bnco} \cite{vesna2022}, 
and to propose future experiments on {\bsco} and {\bcmo}.

\section*{Results}



\noindent {\bf Phonon modes in Ba$_2$CaOsO$_6$}

\noindent First principles density functional theory (DFT) calculations are carried out to explore the phonon
dispersion and structures of 
{\bnco} and {\bsco}. Our motivation here is two-fold: to examine the intrinsic stability of the cubic 
$Fm\bar{3}m$ phase of Ba$_2$CaOsO$_6$, and to explore how an impurity induces local strain fields. 
We study phonon properties within the formulation of density functional perturbation theory (DFPT)
as implemented in the Vienna Ab-initio Simulation Package (VASP) \cite{vasp1,vasp2}. Total energy
and force calculations for structure optimization are carried out using VASP in the plane wave 
pseudo-potential basis, and choosing the generalized gradient approximation (GGA) to describe the 
exchange-correlation functional, and incorporating the effects of Hubbard $U$ and
spin-orbit coupling (GGA+$U$+SOC) on Os (see Methods).

\begin{figure}[t]
\centering
    \includegraphics[width=0.48\textwidth]{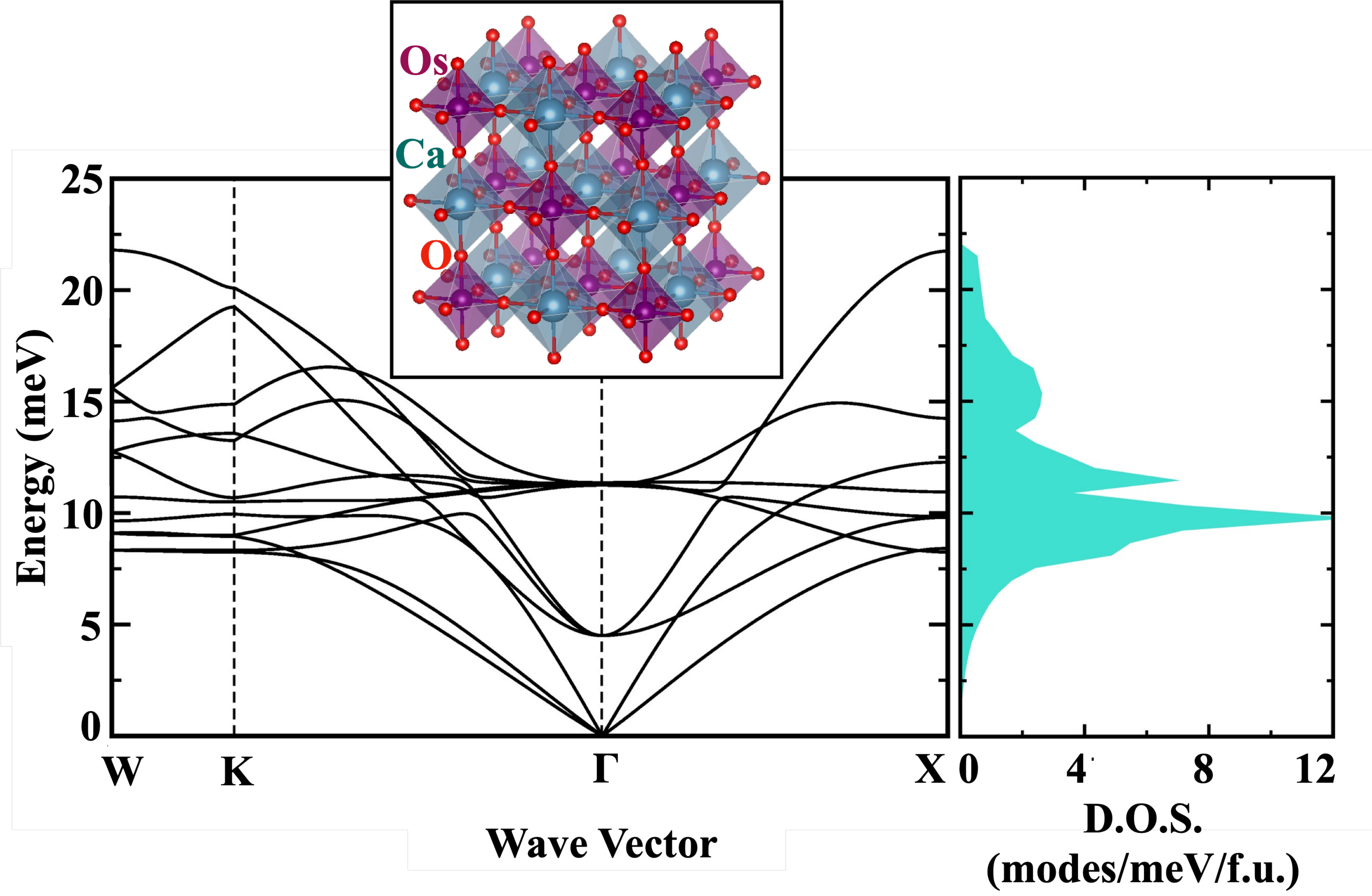}
    \caption{{\bf Phonon dispersion.} {\it Ab initio} results for the 
    low energy phonon dispersion of Ba$_2$CaOsO$_6$ shown along high symmetry paths in the BZ
    of the primitive unit cell. All modes have positive energy reflecting the 
    stability of the cubic $Fm\bar{3}m$ structure.
    Right panel shows the corresponding phonon density of states (DOS).
    Inset: Crystal structure of Ba$_2$CaOsO$_6$, with shaded purple and cyan octahedra depicting
    the checkerboard arrangement of OsO$_6$ and CaO$_6$ units, and red balls indicating O atoms. }
    \label{fig:dft_fig1}
\end{figure}

The computed phonon dispersion of Ba$_2$CaOsO$_6$ in its $Fm\bar{3}m$ structure, plotted along the high-symmetry points of the 
Brillouin zone (BZ) of its primitive unit cell, is shown in Fig.~\ref{fig:dft_fig1} for low energy $E \!<\! 25$\,meV. We find three acoustic modes and nine optical branches in this energy window; as seen from the right panel of
Fig.~\ref{fig:dft_fig1}, this leads to a significant peak in the phonon density of states at 
$E \!\approx \! 10$\,meV, consistent with inelastic neutron scattering results at large momentum transfer \cite{maharaj2019octupolar}. 
We observe no negative phonon frequencies, which thus confirms the intrinsic dynamical stability of the 
$Fm\bar{3}m$ structure with respect to any local non-cubic distortions.
This shows that although the Os is in a $d^2$ configuration, with partially filled $t_{2g}$ orbitals, 
it can be stable against Jahn-Teller distortions from orbital-lattice couplings \cite{Streltsov2020}. 
The absence of any non-cubic distortions
lends support to the scenario that a low energy non-Kramers doublet is realized in the cubic structure of 
Ba$_2$CaOsO$_6$ \cite{maharaj2019octupolar,paramekanti2019octupolar,voleti2020}. The irreducible
representations of the phonon modes at the $\Gamma$-point, as well as phonon spectra extending to higher energies,
are given in Supplementary Note 1.

\bigskip
\noindent {\bf Local strains induced by impurities}

\noindent Having confirmed the stability of the cubic $Fm\bar{3}m$ structure in Ba$_2$CaOsO$_6$, we next 
consider the influence of substitutional impurities like Na or Sr in place of Ca in the structure. 
Since the ionic radius of Na$^{+}$ and Sr$^{2+}$ in octahedral coordination of DP structure are 
$1.02${\AA} and $1.18${\AA}, as compared to $1.00${\AA} for Ca$^{2+}$,
this substitution is expected to cause a local lattice distortion and a
lowering of the local point group symmetry. This propensity to distortion is
signalled by an optical phonon instability in the $Fm\bar{3}m$ structure (see Supplementary Note 1).

\begin{figure*}[t]
\centering
    \includegraphics[width=\textwidth]{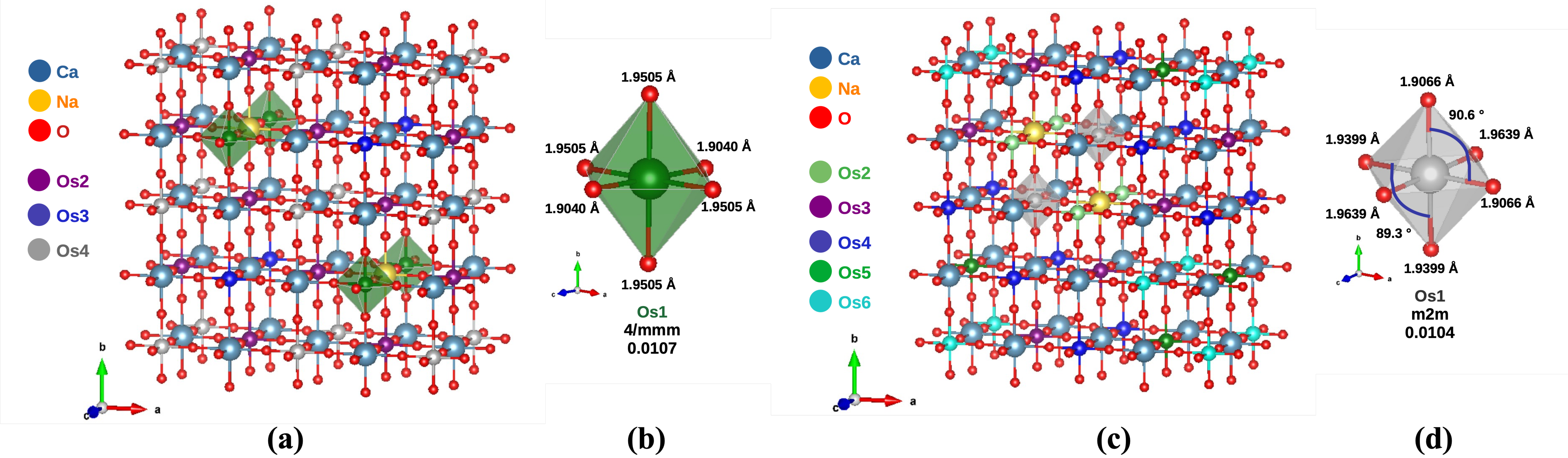}
    \caption{{\bf Optimized structures showing impurity-induced local strains.} 
    {\bf a.} Optimized {\bnco} structure in a supercell (with $16$ formula units) corresponding to 
     $\delta\!=\!1/8$, having two impurity Na atoms substituting for Ca
    at farthest sites. Deep cyan, yellow and red colored balls represent Ca, Na and O atoms, respectively.
    The optimized structure contains four inequivalent Os sites (colored differently). The Os sites
    exhibiting maximum distortion (Os1) are highlighted using shaded green octahedra.
    {\bf b.} Detailed view of the maximally distorted Os1O$_6$ 
    octahedron which have $90^\circ$ O-Os-O bond angles, but with variations in the bond length, together with
    the local site-symmetry ($4/mmm$), and $D$ parameter ($D=0.0107$) 
    which measures the degree of octahedral distortion (see text for details).
    This leads to an effective $E_g$ strain field on the Os1 site.
    {\bf c.} Same as {\bf a}, but with two Na atoms substituting for nearest pairs of Ca. In this case six different 
    inequivalent Os sites are created and depicted with different colored balls; Os sites
    exhibiting maximum distortion Os1 are highlighted with grey octahedra.
    {\bf d.} Same as {\bf b} but shown for maximally distorted Os1 site in near configuration (see text for details).
     In this case, the site symmetry is lowered to $m2m$, $D=0.0104$, and O-Os-O bond angles deviate from $90^\circ$ which leads
     to an additional $T_{2g}$ strain field on Os1. Similar results are found for {\bsco}.}
        \label{fig:dft_fig2}
\end{figure*}

To investigate, in more detail, the effect of substitutional impurities and impurity clustering, we next study a {\bco} supercell 
of dimension 2$\times$2$\times$1 in terms of a cubic four formula unit cell. This results in $16$ Ca sites in the supercell.
We replace two of these Ca by Na or Sr, amounting to an effective impurity concentration 
$\delta\! =\! 1/8$ for {\bnco} and {\bsco} respectively.
We consider two configurations in the supercell: (i) the far configuration shown in Fig.~\ref{fig:dft_fig2}{\bf a} 
and (ii) the near configuration shown in Fig.~\ref{fig:dft_fig2}{\bf c}.

In the far configuration, shown in Fig.~\ref{fig:dft_fig2}{\bf a}, we substitute two distant Ca atoms in the supercell with
the substitutional impurity.
Full structure optimization of this configuration leads to four inequivalent Os sites marked by different colors and labelled
Os1-Os4, in Fig.~\ref{fig:dft_fig2}{\bf a} and \ref{fig:dft_fig2}{\bf b} (see Supplementary Notes 1 and 2 for details).
Of these four inequivalent Os sites, the maximal distortion occurs at sites which have
two Na atoms at its nearest neighbor shell (depicted as green octahedra). 
The site symmetry of this Os site, marked Os1 and shown in Fig.~\ref{fig:dft_fig2}{\bf b}, 
is found to be $4/mmm$. 
As discussed below, this leads to an $E_g$ strain on the Os1 $d$-orbitals.
The distortion of OsO$_6$ octahedra can be measured by the distortion index, 
$D = \frac{1}{6} \sum_{i=1}^{6} \frac{\vert l_i - l_{av}\vert}{l_{av}}$ where $l_i$ is the 
bond-length of the $i$-th Os-O bond, and $l_{av}$ is the average bond length; for 
Os1O$_6$ we find $D\!\approx\!0.01$. We show in the Supplementary Note 5 that this 
$1$-$2$\% $E_g$ strain translates to a $5$-$10$\,meV splitting of the non-Kramers doublet.

We next turn to the near configuration shown in Fig.~\ref{fig:dft_fig2}{\bf c}, with
two Na atoms substituting for neighboring Ca atoms. In this case,
full structure optimization leads to six inequivalent Os sites marked Os1-Os6 (for details see 
Supplementary Note 2).
We find that the maximum distortion of Os-O bond lengths and bond angles 
occurs at the site marked Os1 which is adjacent to both Na atoms, with its site symmetry being lowered from 
$4/mmm$ to $m2m$ (depicted as grey octahedra), and $D\!\approx\!0.01$.
As discussed below and in Supplementary Note 2, this leads to an additional $T_{2g}$ strain at this Os site.
Finally, the distortion index $D$ for sites further from the impurity site 
are an order of magnitude smaller (see Supplementary Note 2). As discussed later,
the presence of this $T_{2g}$ strain is crucial for theoretically modelling the
experimentally observed NMR spectra.

\bigskip

\noindent {\bf Landau theory of multipole-strain coupling}

\noindent Our next goal is to understand how the impurity induced strain, discovered in our
{\it ab initio} structure calculations, can impact multipolar orders. 
We resort here to a symmetry based approach, beginning with a brief
review of the multipoles and their symmetries for a $d^2$ electron 
configuration. We next turn to a Landau theory for the coupling the
multipoles to strain fields, and discuss a microscopic basis 
for this Landau theory where strain fields are shown to arise from impurity
induced deformations.


For two electrons in the $t_{2g}$ orbital, the naively
expected ground state is a $J\!=\!2$ moment,
which we can view as arising from spin-orbit locking of
a total orbital angular momentum $L=1$ and total spin $S=1$. This
gets split into a non-Kramers ground state doublet, and
an excited
triplet with a gap $\Delta \!\sim\! 10$-$20$\,meV \cite{voleti2020},
in reasonable agreement with the spin gap observed in
inelastic neutron scattering experiments \cite{maharaj2019octupolar}
(see Supplementary Note 3).
Working in the $| J_z = m\rangle$ basis, the non-Kramers doublet 
wavefunctions are given by
\bea
|\psi_{g,\uparrow}\rangle =  \frac{1}{\sqrt{2}} (|2\rangle + | -2 \rangle);~~~
|\psi_{g,\downarrow}\rangle = |0\rangle.
\eea
Within this non-Kramers doublet space, the Pauli matrices $\tau_x,\tau_y,\tau_z$ are proportional to multipole operators, 
and are given by
$\tau_x \!\equiv\! (J_x^2\!-\!J_y^2)/2\sqrt{3}$,
$\tau_y \! \equiv\! \overline{J_x J_y J_z}/6\sqrt{3}$,
and $\tau_z \!\equiv\! (3 J_z^2\!-\! J(J+1))/6$, with overline denoting symmetrization.
Here, $\tau_x,\tau_z$
are electric quadrupoles while $\tau_y$ is a magnetic octupole.
The ground state doublet manifold has vanishing matrix elements for the dipole operators $\vec J$, precluding magnetic dipole ordering.
However, the ground state doublet can
lead to broken time-reversal symmetry below $T_c$ (while preserving cubic symmetry)
if $\langle \tau_y \rangle \neq 0$ which corresponds to ferro-octupolar ordering.

To examine the impact of strain, we note that the 
non-Kramers doublet degeneracy is protected by the crystalline point group symmetry. We thus
expect that (random) local strains will break this degeneracy. Such strains may be produced
by impurities, which can locally suppress the octupolar order while favoring quadrupolar orders.
Landau theory provides a useful symmetry-based approach to study the coupling of strain to
multipoles \cite{Patri2019}.
To explore this here, we begin by considering time-reversal invariant
perturbations acting in the $J=2$ space, focussing on 
the simplest $E_g$ and $T_{2g}$ strain fields.
When we later discuss the microscopic basis for this Landau theory, 
our focus will be on strain fields localized at Os sites directly adjacent to the impurity site. 
While our previously discussed {\it ab initio} calculations yield a more complex 
distortion pattern of OsO$_6$ octahedra even far away from the impurity site,
those longer range distortions are found to be order of magnitude smaller; this will 
justify our simplified microscopic viewpoint.

The $E_g$ strain perturbations act as fields $(h_x,h_z)$ 
which couple to the low energy quadrupolar degrees of freedom 
$(\tau_x, \tau_z)$ within the non-Kramers doublet. On symmetry grounds,
these couplings take
the form
\begin{eqnarray}
\!\!\!\!\!\!\! \delta H_{x} &=& - d_1 \!\! \int\! d^3\br (\varepsilon_{xx}(\br)\!-\! \varepsilon_{yy}(\br)) \tau_x(\br) \\
\!\!\!\!\!\!\! \delta H_{z} &=& - d_2 \!\! \int\! d^3\br \frac{1}{\sqrt{3}}
(2 \varepsilon_{zz}(\br)\! -\! \varepsilon_{xx}(\br)\!-\! \varepsilon_{yy}(\br)) \tau_z(\br)
\end{eqnarray}
We expect that these $E_g$ strain fields can suppress the octupolar
transition temperature since they act as transverse fields on the
Ising octupolar order. We confirm this later using Monte Carlo simulations. For a sufficiently
strong local strain field, it may be possible to locally kill the octupolar
order.
The strain induced local quadrupolar order 
can lead to broadening of NMR
spectra via generation of electric field gradients. 

By contrast, the $T_{2g}$ strain tensor components, such as
$\varepsilon_{xy}(\br)$, lead to a coupling between the low energy non-Kramers doublet and 
the excited magnetic triplet. This set of Hamiltonian perturbations take the form
\begin{eqnarray}
\delta H_{\rm od} &=& - d_3 \int d^3\br  
\left[\varepsilon_{xy}(\br) Q_{xy}(\br)
+\varepsilon_{yz}(\br) Q_{yz} (\br) \right. \nonumber \\
&+& \left. \varepsilon_{xz}(\br) Q_{xz} (\br)  \right]
\end{eqnarray}
where $Q_{\alpha\beta}=(J_\alpha J_\beta+J_\beta J_\alpha)/2$ 
has no matrix elements in the non-Kramers doublet subspace.
In a octupolar ordered state with nonzero
$\langle \tau_y(\br) \rangle$, taking into account the first-order 
perturbative correction to the wavefunction due to mixing with 
the excited magnetic triplet, these 
couplings induce
weak dipole moments in the ground state. Defining
$\bm{\varepsilon} \equiv (\varepsilon_{yz},\varepsilon_{zx},\varepsilon_{xy})$, we arrive at
\begin{eqnarray}
\langle\vec J (\br) \rangle = \frac{2 d_3}{\Delta}
\bm{\varepsilon}(\br) \langle \tau_y(\br)\rangle
\end{eqnarray}
The $T_{2g}$ strain fields thus reveal the hidden octupolar order 
by generating `parasitic' dipole moments. If the correlation length of 
these strain fields, and thus the parasitic dipole moments 
is short-ranged, these moments may provide a dephasing mechanism for oscillations
in $\mu$SR experiments in the octupolar ordered state.
We note that these parasitic dipole moments 
may also be observable in NMR experiments via their hyperfine coupling to nuclear dipole moments,
an effect we will later model carefully.

\begin{figure*}[t]
    \centering
    \includegraphics[width=\textwidth]{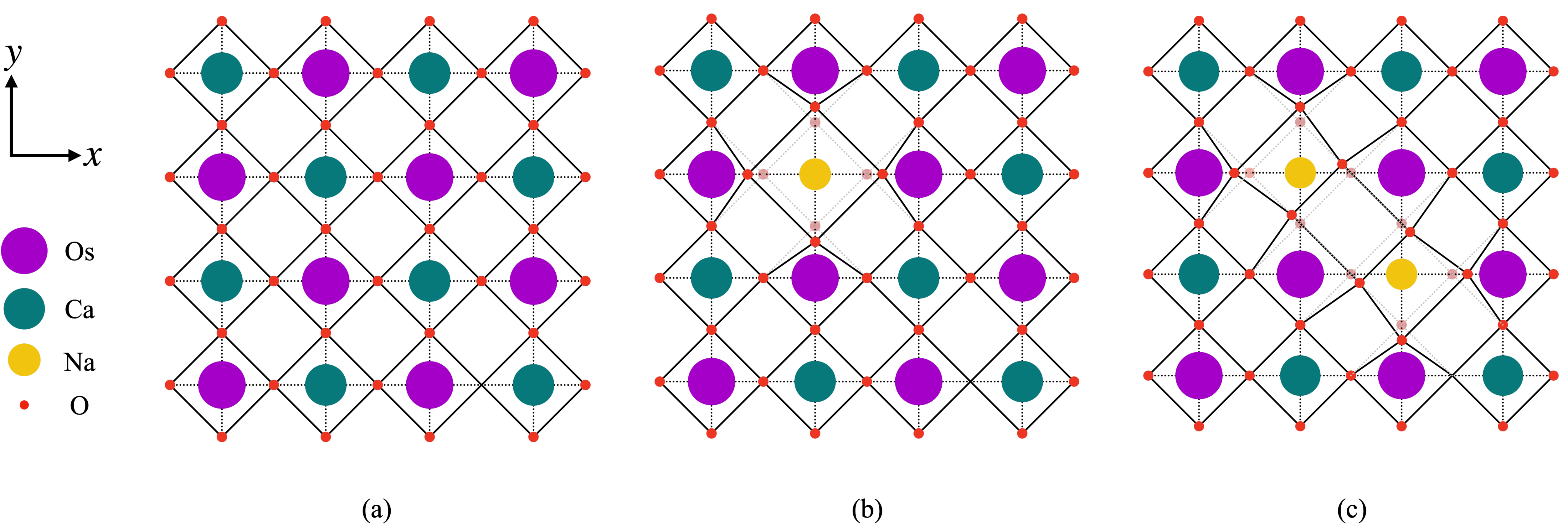}
    \caption{{\bf Schematic strain field configurations.} Simplified schematic crystal structure in single $xy$ plane of the osmate cubic double perovskites. {\bf a.} Ideal crystal structure in the absence of impurities. {\bf b.} $A_{1g}$ breathing distortion around an isolated impurity (Na$^{+}$ or Sr$^{2+}$) site. The light dashed lines and light red dots mark the original positions of oxygens in the undistorted cubic structure. This leads to an effective $E_g$ octahedral distortion acting on $d$-orbitals for the six neighboring Os 
    sites as discussed
    in Supplementary Note 5. (Note that there is an additional pair of Os atoms above and below the impurity, not shown, that are affected in a similar way.) {\bf c.} Modified distortion in the case of a pair of nearby impurities. In this case, the breaking of an additional mirror symmetry (vertical $xz$ or $yz$ plane passing through Os-Na-Os axis) leads to further symmetry lowering. This
    induces effective $T_{2g}$ strain fields on $d$-orbitals on the six neighboring osmiums (see Supplementary Note 5).}
    \label{fig:SchematicDistortion}
\end{figure*}

One way such local strain fields could appear in a material is via a small density
of impurities, such as antisite impurities, oxygen vacancies, or 
impurities induced by intentional chemical substitution. 
Our {\it ab initio} results
show that when a single Ca$^{2+}$ is replaced with Na$^+$ or Sr$^{2+}$, 
the six Os$^{6+}$ ions neighboring the impurity site will have 
their oxygen octahedra get uniaxially distorted from their original equilibrium 
positions.
A schematic picture of the ideal cubic structure and a simplified
model of the strain field in the $xy$ plane induced by a single impurity is shown in
Figs.~\ref{fig:SchematicDistortion} {\bf a} and \ref{fig:SchematicDistortion} {\bf b}. 
The depicted distortions are based on our {\it ab-initio} results, but have been 
simplified, as justified above, to only focus on distortions immediately adjacent to the impurity site.
With Os$^{6+}$ at the center of the distorted
octahedra as the local center of inversion,
such an octahedral distortion can be decomposed into odd-parity and
even-parity modes, with only the latter having matrix elements in the $d$-orbital
space; see Supplementary Note 5. This leads to $E_g$ strain fields
for the $d$-orbitals on the Os$^{6+}$ site. With increasing impurity concentration,
there is a higher probability of neighboring pairs of Na$^+$ 
or Sr$^{2+}$ impurities. In this case, the local symmetry around Os$^{6+}$ ions is 
further reduced, as found from our {\it ab initio} calculations, and
shown in Fig.~\ref{fig:SchematicDistortion}{\bf c}. The strain fields at the
Os$^{6+}$ site then also include a $T_{2g}$ component; we derive this in Supplementary Note 5. 
Thus, impurities induce both $E_g$ and $T_{2g}$ strain fields, providing a firm microscopic
basis for our Landau theory.

\bigskip

\noindent {\bf Impurity-induced $T_c$ suppression}


\noindent To study how impurities suppresses the octupolar $T_c$, we have carried out MC simulations for the 
pseudospin Hamiltonian $H_0 + H_{\rm imp}$ describing Os$^{6+}$ non-Kramers doublets 
on the face centered cubic (fcc) lattice.
Here $H_0$ is the nearest-neighbor pseudospin Hamiltonian in the absence of impurities 
which has been previously obtained
using microscopic calculations \cite{Voleti2021,Churchill2022}, and $H_{\rm imp}$ incorporates
the transverse fields adjacent to the randomly located impurity sites. We note that
such classical MC simulations for a spin-$1/2$ system, while not capturing the effects of
quantum fluctuations, can nevertheless provide a useful guide to the physics, and
has been used to explore 2D Kitaev materials \cite{KitaevMC_Perkins_PRL2012,KitaevMC_Janssen_PRL2016} 
and 3D spin-ice compounds \cite{spinice_Changlani_npjQM2022}.

The most general symmetry allowed $H_0$ is given by
\begin{eqnarray}
\!\! H_{0} \!&=&\!\! \sum_{\langle i,j\rangle}\! \left[ K_{\rm o} \tau_{i y} \tau_{j y}
\!+\! \left( K_1 \cos^2\!\phi_{ij} \!+\! K_2 \sin^2\!\phi_{ij} \right) \tau_{i x} \tau_{j x} \right. \nonumber \\ 
 &+& \left( K_1 - K_2  \right) \sin\phi_{ij} \cos\phi_{ij} \left( \tau_{ix}\tau_{jz} + \tau_{iz}\tau_{jx} \right) \nonumber \\ 
 &+&\left.  \left( K_1 \sin^2\phi_{ij} + K_2 \cos^2\phi_{ij}  \right)\tau_{i z} \tau_{j z}\right]
\end{eqnarray}
where $\phi_{ij} = \{ 0 , 2\pi/3 , 4\pi/3 \}$ correspond to nearest neighbors $(i,j)$ in the $\{ xy , yz , zx \}$ planes.
$K_{\rm o}$ and $K_{1,2}$ respectively correspond to the octupolar exchange and quadrupolar couplings. We consider two
sets of model exchange couplings as shown in Table \ref{table:1}. While Model A has exchange couplings motivated
by our previous study \cite{Voleti2021}, Model B explores the case of significantly enhanced quadrupolar exchange 
interactions.

We next incorporate the effect of impurities.
When we replace a fraction $\delta$ of
Ca$^{2+}$ by Sr$^{2+}$ or Na$^{+}$, the six Os$^{6+}$ neighbors of the substituted site
experience an $E_g$ octahedral distortion as discussed above. This leads to local quadrupolar fields 
which couple as
\begin{equation} \label{strainfieldeq}
    H_{\rm imp} = \sum_{i,\alpha=x,z} \!\! \vec h^\alpha_{\rm imp}(i) ~\tau_{i\alpha}
\end{equation}
where the quadrupolar fields $\vec h_{\rm imp}(i) \!\neq\! 0$ {\it only} for sites adjacent to
an impurity. These fields point 
along distinct directions for 
the Os$^{6+}$ neighbors of the impurity site:
\begin{eqnarray} 
\pm x~{\rm neighbors}&:&~\vec h_{\rm imp} = h ( \sqrt{3} \hat{x} - \hat{z})/2 \\
\pm y~{\rm neighbors}&:&~\vec h_{\rm imp} = h (-\sqrt{3} \hat{x} - \hat{z})/2 \\
\pm z~{\rm neighbors}&:&~\vec h_{\rm imp} = h \hat{z} \label{streqz}
\end{eqnarray}
The field strength $h$ is
proportional to the distortion induced by the impurity. If there are adjacent impurities, we add up the
corresponding quadrupolar fields on the neighboring Os$^{6+}$ sites to get the total $\vec h_{\rm imp}(i)$. 
These quadrupolar fields act as transverse fields on the Ising octupolar order.
We note that the {\it transverse field strength and orientations are not random} - however, the locations in space 
where they act is dictated by the random impurity positions.
Using the $\sim\!1$-$2\%$ strain inferred from our {\it ab initio} calculations,
we estimate $h$ to be on the order of $\sim\!5$-$10$\,meV; see Supplemetary Note 5 for details on this estimate of $h$.
Below we will explore a range of values in our MC simulations.

We ignore the $T_{2g}$ strains for the MC simulations; 
since these strains do not act directly within the non-Kramers doublet subspace, their impact on octupolar ordering
is weaker. While $T_{2g}$ strain field induce parasitic dipole moments to first order in the strain (in the octupolar broken
symmetry phase), 
they modify the pseudospin Hamiltonian itself only at second order in strain.

\begin{figure}[t]
\centering
    \includegraphics[scale=0.5]{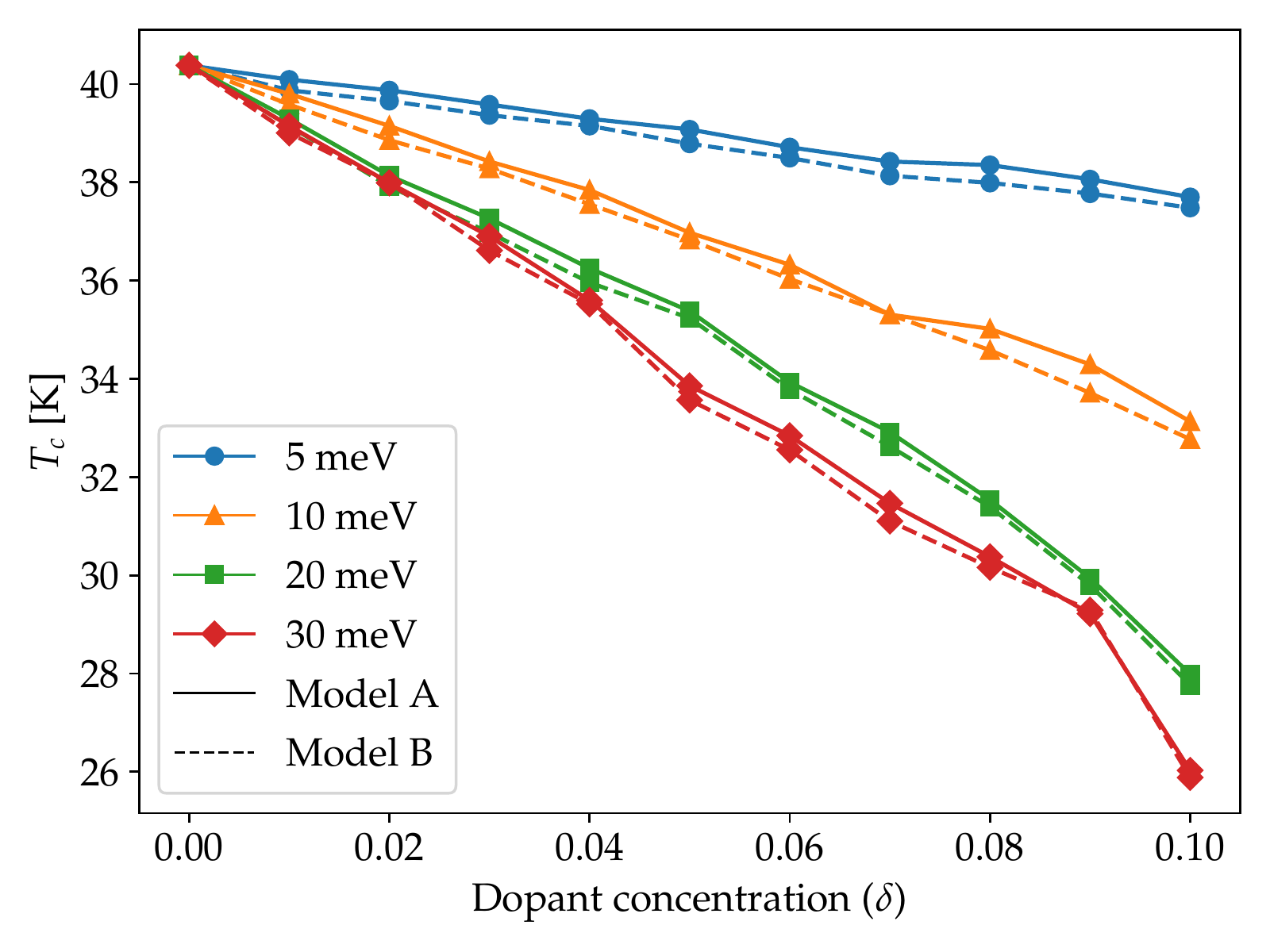}
    \caption{{\bf Impurity induced $T_c$ suppression.} 
    Evolution of the ferro-octupolar $T_c$ with increasing impurity concentration (at the
    Ca$^{2+}$ site) for different values of the distortion-induced quadrupolar field $h$ induced on
    neighboring Os$^{6+}$ sites. We show results for both model $A$ (solid line) and model $B$ (dashed
    line) with exchange
    parameters listed in Table \ref{table:1}.}
        \label{fig:Tc_versus_X}
\end{figure}

We have carried out MC simulations of the Hamiltonian $H_0 + H_{\rm imp}$ for 
different impurity concentrations and a range of local quadrupolar pinning field strengths 
$h$ (see Methods).
At zero impurity concentration ($\delta\!=\!0$), our model
yields a ferro-octupolar transition temperature $T_c \! \approx\! 40$\,K, comparable with
experiments on the undoped osmate double perovskites \cite{Thompson_JPCM2014,Kermarrec2015,Thompson_PRB2016,MarjerrisonPRB2016,maharaj2019octupolar}.
As seen from Fig.~\ref{fig:Tc_versus_X}, 
the octupolar $T_c$ decreases nearly linearly with increasing impurity concentration $\delta$.
For larger $h$,the slope $(dT_c/d\delta)$ increases but eventually becomes independent of $h$ 
since the local quadrupolar order gets perfectly
pinned at large $h$. 
The exchange parameter sets for model $A$ and model $B$ are found to lead to similar results, with no
significant differences either in the octupolar $T_c$ or in the evolution of $T_c$ with impurity concentration.

\begin{figure}[t]
\centering
    \includegraphics[scale=0.5]{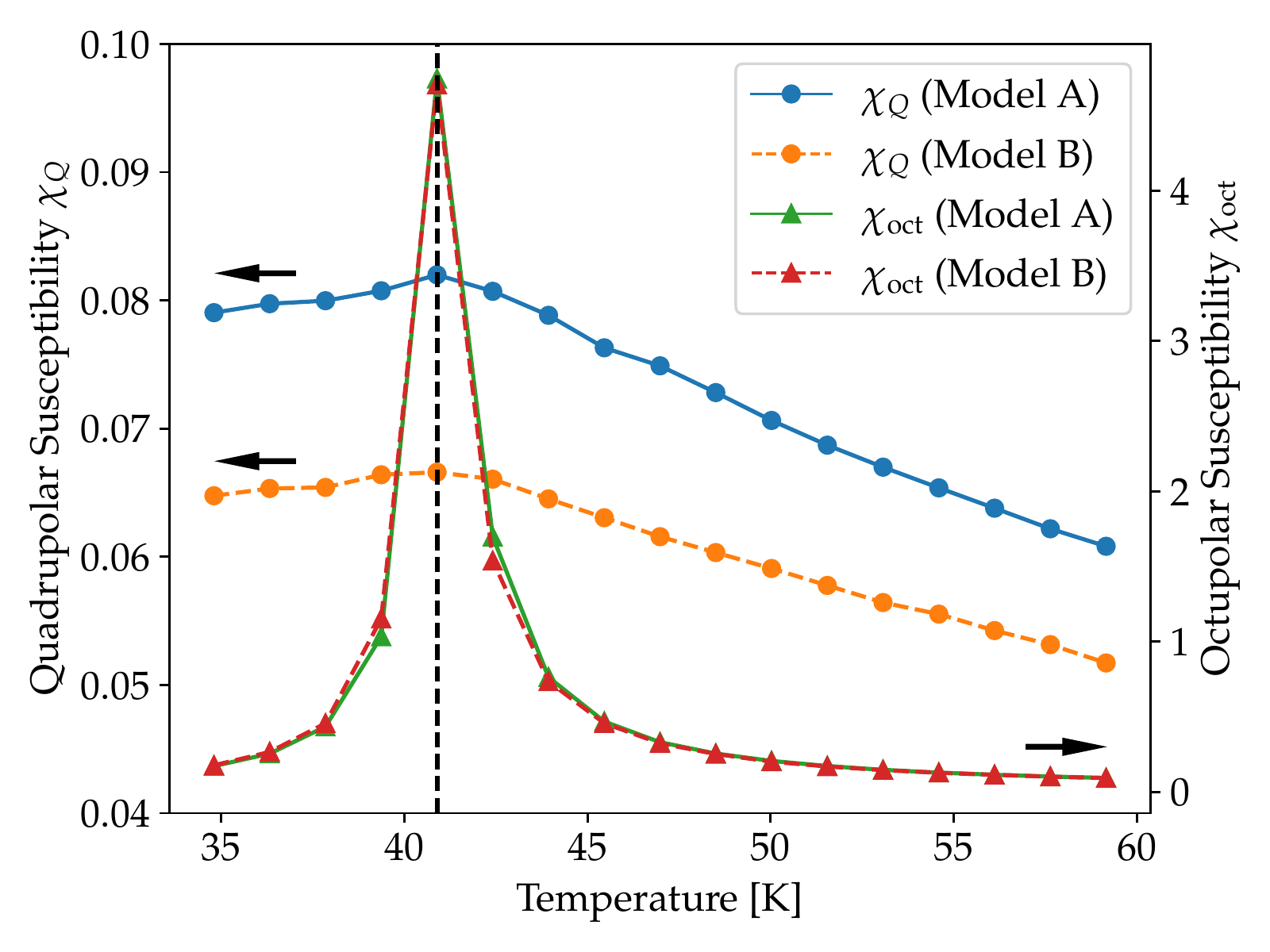}
    \caption[]
        {{\bf Octupolar and quadrupolar susceptibilities.} MC results in the clean case for 
        (i) the octupolar $\chi_{\rm Oct}(T)$ susceptibility which exhibits a divergence at the ferro-octupolar $T_c$,
        and (ii) the uniform  quadrupolar $\chi_Q(T)$ susceptibility which has a Curie-Weiss form at high temperature but
        gets cut off at $T_c$ and drops as we further cool into the ferro-octupolar symmetry broken phase. Results are
        shown for both model exchange parameters from Table \ref{table:1}.}
        \label{fig:Chi}
\end{figure}

The impact of impurities on local
quadrupolar order may also be indirectly deduced from the components of the
multipolar susceptibility tensor 
$\chi_{a b}(T)$ in the clean $\delta=0$ limit. The uniform susceptibility is given by
\begin{eqnarray*}
\chi_{ab}(T)={\beta} \left( \left\langle M_a M_b\right\rangle - \left\langle M_a \right\rangle \left\langle M_b 
\right\rangle \right)
\end{eqnarray*}
where $a,b\! \in\! (x,y,z)$, and $M_a\!=\! \sum_i \tau_i^a /2$.
Fig.~\ref{fig:Chi} shows the temperature evolution of the octupolar $\chi_{\rm oct}(T) \!\equiv\! \chi_{yy}$
and quadrupolar susceptibility $\chi_{Q}(T)\!\equiv\!\chi_{xx}\!=\!\chi_{zz}$.
The octupolar susceptibility diverges at the ferro-octupolar $T_c$. By contrast the quadrupolar
components exhibit a Curie-Weiss type behavior at high temperature, but this growth eventually
get cutoff at $T_c$ and it decreases slightly due to the onset of octupolar order. Model $B$ with
a larger antiferro-quadrupolar exchanges leads to a smaller uniform $\chi_{Q}(T)$.
In the presence of a
strain induced perturbation $h$, we expect the induced
local Os quadrupole moments in the vicinity of the impurity to qualitatively track the temperature
dependence of $\chi_{Q}(T)$.
This will play a role below in our understanding of the impurity NMR spectra.

Thermodynamic and $\mu$SR measurements on the osmate DPs Ba$_2$CaOsO$_6$, Ba$_2$MgOsO$_6$, and Ba$_2$ZnOsO$_6$
have found evidence for a single phase transition which does not involve a change in crystal symmetry but at which 
time-reversal symmetry is broken \cite{Thompson_JPCM2014,Kermarrec2015,Thompson_PRB2016,MarjerrisonPRB2016,maharaj2019octupolar}. 
These results are consistent with a ferro-octupolar ordering transition.
Recent $\mu$SR and NMR results on {\bnco} \cite{vesna2022} 
show that $T_c$ gets nearly linearly suppressed with increasing impurity
concentration $\delta$, as we go away from the clean limit $\delta=0$. Our model results in 
Fig.~\ref{fig:Tc_versus_X}
showing the change of $T_c$ with impurity concentration are
consistent with these experimental results. For $h\!=\!10$\,meV, we find
$(1/T_c) (dT_c/d\delta) \! \approx\! -2$, which is in agreement with experiments on {\bnco} which
find $(1/T_c) (dT_c/d\delta) \! \approx\! -2.2$ \cite{vesna2022}. With increasing impurity concentration,
this model is expected to yield a disordered quantum critical point (QCP) where the octupolar $T_c$ vanishes.
Naively extrapolating our low doping results, this QCP is expected to occur at large dopant 
concentrations $\delta\! \gtrsim \! 0.5$; however, quantum fluctuations may shift
this QCP to smaller dopant concentrations.
\bigskip

\noindent {\bf NMR spectra}

\noindent We next turn to modelling the $^{23}$Na NMR spectra. Since the $^{23}$Na nucleus has
spin $I=3/2$, we consider symmetry allowed terms for the interaction of this nuclear spin with the applied
Zeeman field and the neighboring Os multipole moments:
\begin{align} \label{localnmr}
H_{\rm Na} =& - \sum_i \gamma(i) ~\hat{\vec I}(i) \cdot \vec B 
+ g_1 \sum_{\langle i,j \rangle}  \hat{\vec I}(i) \cdot \tilde{\pmb{\varepsilon}}_j \langle \tau^y_{j} \rangle \nonumber\\
&+ \sum_{\langle i,j \rangle} \hat{\bf Q}^{\rm Na}(i) \cdot \mathbb{A}_{ij} \cdot \langle {\bm{\mathcal{Q}}_j} \rangle 
\end{align}
Here, $\gamma(i)$ is the gyromagnetic ratio (units: MHz/Tesla), which can be renormalized by the $A_{1g}$ distortion
around the Na site leading to a Knight shift. We define it as
\begin{equation}
\gamma(i) = \gamma^{\rm Na} \left( 1 + g_0 \sum_{j \in {\rm nbr}(i)} \langle \mathcal{Q}_j^{A_{1g}} \rangle \right)
\end{equation}
where $\gamma^{\rm Na}\!\approx\!11.262$ MHz/T is the bare gyromagnetic ratio, 
and the $A_{1g}$ distortion on the Na site is
constructed as a sum of specific quadrupole moments
the Os neighbors $j$, namely:
\begin{equation}
    \langle \mathcal{Q}_j^{A_{1g}} \rangle = 
    \begin{dcases}
        {\sqrt{3} \over 2} \la \tau_j^x \ra - {1 \over 2}\la \tau_j^z \ra  & \text{for } \langle ij \rangle \text{ along } \hat{x} \\
        -{\sqrt{3} \over 2} \la \tau_j^x \ra - {1 \over 2}\la \tau_j^z \ra  & \text{for } \langle ij \rangle \text{ along } \hat{y} \\
        \langle \tau_j^z \rangle & \text{for } \langle ij \rangle \text{ along } \hat{z}. \\
    \end{dcases}
\end{equation}
For an isolated Na$^{+}$ ion, this sum over the six neighbors preserves the octahedral point group symmetry at the Na site, 
and thus it only 
leads to a temperature-dependent renormalization of the Knight shift.

The second term in Eq. \eqref{localnmr} is the effect of the dipole moments on the Os atoms generated via $T_{2g}$ distortions, as explained previously. This term only contributes when there are two neighbouring impurity Na ions, causing $T_{2g}$ distortions on two of the Os atoms which are adjacent to both Na. We assume an Ising-like interaction between the parasitic dipole component produced by the $T_{2g}$ strain and the 
dipole moment of the Na; here, $\tilde{\pmb{\varepsilon}}_j$ is a unit vector obtained by normalizing the $T_{2g}$ distortion 
vector ${\pmb{\varepsilon}}_j$ defined earlier. For instance, for a nearby pair of Na in the $xy$
plane, $\tilde{\pmb{\varepsilon}}_j = \pm \hat{z}$ (see Supplementary Note 5). 
The magnitude of the $T_{2g}$
distortion, which is not known in detail, is absorbed into the coupling constant $d_1$.

The third term accounts for the quadrupolar couplings between the Na and Os atoms. Since the Os atom only has $E_g$ quadrupolar moments in the low energy doublet, we only consider, for simplicity, interactions between those and $E_g$ quadrupoles of the Na nuclear spin.
The local operators in this term are defined as 
\begin{equation}
    \hat{\mathbf{Q}}^{\rm Na} = \begin{pmatrix}
       \hat{I}_x^2 - \hat{I}_y^2 \\
        \left( 3 \hat{I}_z^2 - I(I+1) \right) / \sqrt{3}
    \end{pmatrix} 
\end{equation}
and 
\begin{equation}
    \la \pmb{\mathcal{Q}} \ra = \begin{pmatrix}
        \la \tau_x\ra \\ \la \tau_z \ra
    \end{pmatrix}
\end{equation}
Along the $\hat{x}$ bonds, we assume a diagonal interaction, i.e.
\begin{equation}
    \mathbb{A}_{ij} = \begin{pmatrix}
        a_1 & 0 \\
        0 & a_2 
    \end{pmatrix} \quad \text{for } \langle ij \rangle \text{ along } \hat{x}
\end{equation}
We can then use rotations to infer the corresponding transformed $\mathbb{A}$ matrix for Na-Os bonds
along the $\hat{y}$ and $\hat{z}$ directions. For simplicity, we set $a_1\!=\!a_2\!=\!a$.


We choose to measure the various couplings used in the Hamiltonian $H_{\rm Na}$ in Eq.~\ref{localnmr} 
in units of $\gamma^{\rm Na} B \equiv \omega_0$. For a typical field $B=10$T, we get $\omega_0 \!\approx\! 
112.6$MHz. The couplings we use for illustrative NMR plots are
given by $(g_0,g_1,a) = (1,5,-1) \times 10^{-3}~\omega_0$. This choice of couplings
is found to provide a reasonable semi-quantitative description of the experimental NMR data \cite{vesna2022}; however, the
key spectral features discussed below are not extremely sensitive, at a qualitative level, to the precise values
of these couplings.
Previous analyses of such spectra often focus only on the average quadrupole-quadrupole couplings.
This misses the subtle effects of the remaining terms in 
Eq. \eqref{localnmr}, specifically the couplings $(g_0,g_1)$.

\begin{figure*}[t]
\centering
    \includegraphics[width=0.75\textwidth]{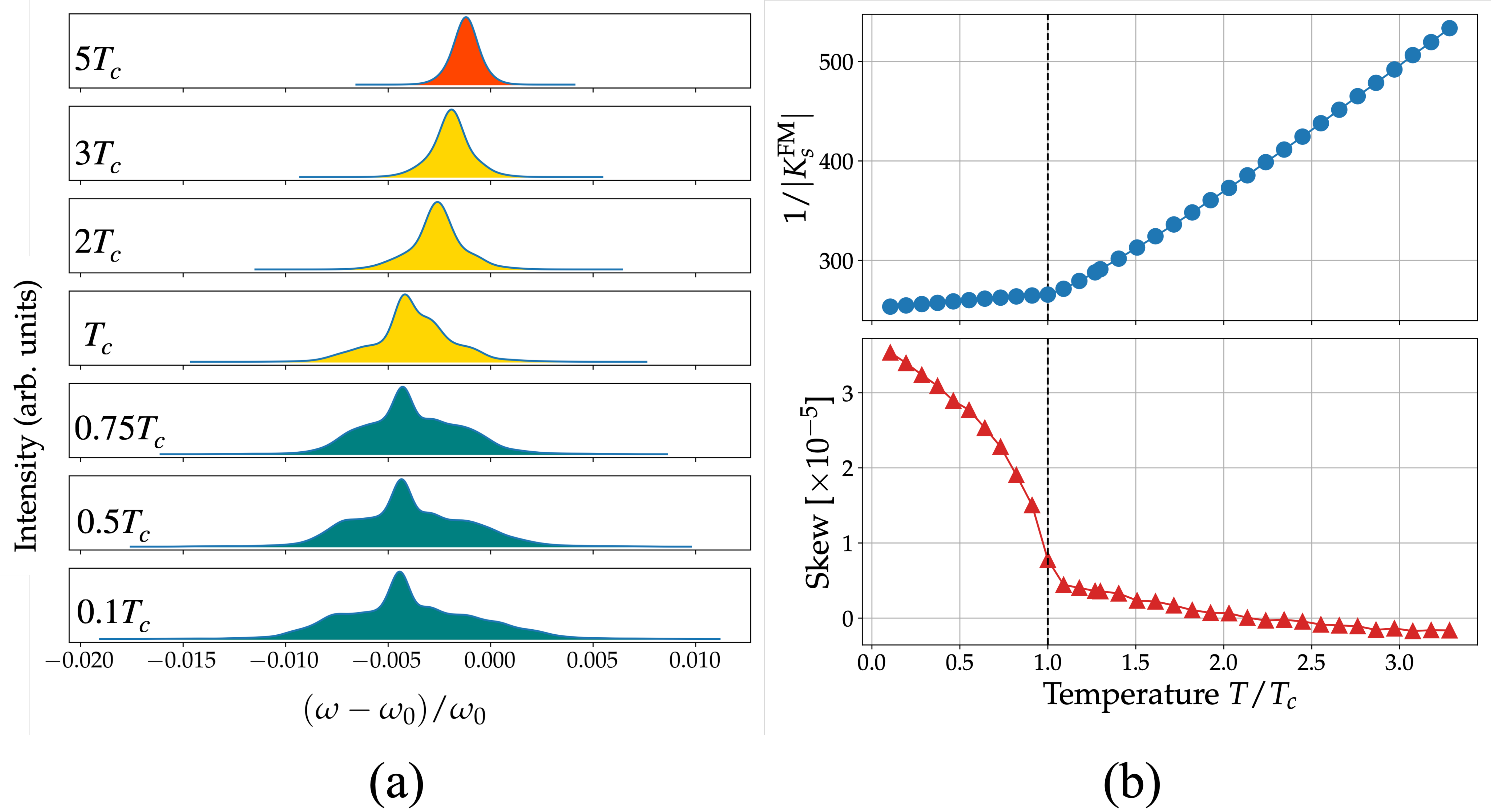}
    \caption[]
        {{\bf Temperature dependent NMR spectra.} {\bf a.} Computed $^{23}$Na NMR spectra over a wide temperature range. 
        At high temperature (red: $T=5 T_c$), there is a narrow peak due to motional 
        narrowing as we sample many fluctuating Os moment configurations. The shift of this peak upon cooling reflects the
        growth of quadrupolar susceptibility, while its broadening and asymmetry arises from moments getting locally pinned at low
        symmetry sites (yellow: $T=3 T_c, 2 T_c, T_c$). 
        The additional extremely large broadening below $T_c$ (blue: $T=0.75 T_c, 0.5 T_c, 0.1 T_c$)
        reflects the formation of inhomogeneous parasitic dipole moments from the interplay of strain and octupolar order.
        {\bf b.} Top: Inverse Knight shift obtained from the first moment $\langle \omega \rangle$ of the theoretical
        NMR spectrum as $1/K^{\rm FM}_s\!=\! \omega_0/(\langle \omega \rangle-\omega_0)$. Bottom: Skew of the NMR
        spectrum showing the growth of spectral asymmetry upon cooling, with a singular increase below $T_c$ which arises
        due to coupling of the $^{23}$Na octupole to the ordered Os ferro-octupolar moment $\langle \tau_y \rangle$.}
        \label{fig:nmr}
\end{figure*}

Fig.~\ref{fig:nmr} presents our results for the $^{23}$Na NMR spectrum where we
have averaged over external Zeeman field orientations as appropriate for a powder sample (see Methods and Supplemtary Note 6 for
details).
These results are for an impurity concentration $\delta=0.1$
which is the smallest value of Na concentration which has been experimentally explored \cite{vesna2022}.
This puts us close to the clean end point {\bco}, so our starting point of assuming a 
dilute concentration of impurities is a valid approximation. 

Fig.~\ref{fig:nmr}{\bf a} shows the calculated NMR spectra (see Methods) over a range of temperatures, 
ranging from high temperatures $T=5 T_c$ to low
temperatures within the octupolar ordered state $T=0.1 T_c$. Fig.~\ref{fig:nmr}{\bf b} 
shows the inverse Knight shift (associated with the first moment) and the skew of the spectrum as a function of temperature. We
describe below key features of the theoretically calculated NMR spectra which capture the
experimental observations \cite{vesna2022} remarkably well.

\noindent i) At high temperature, e.g. $T=5 T_c$ in Fig.~\ref{fig:nmr}{\bf a}, the NMR spectrum exhibits
a single narrow peak consistent with motional narrowing since the effect of the thermally fluctuating neighbor
Os multipole moments average to zero. This single peak shifts upon cooling, with the inverse Knight shift ($1/K_s$)
scaling approximately linearly with $T$. The systematic temperature dependence of 
the first moment of the spectrum is shown in Fig.~\ref{fig:nmr}{\bf b}.
We attribute this observed temperature dependence to the temperature-dependent quadrupolar susceptibility $\chi_Q(T)$ of Os
shown in Fig.~\ref{fig:Chi}. Our scenario is that
the impurity induces quadrupolar moments on the six neighboring Os atoms, which scale as $\sim \chi_Q(T) h$, where $h$ is
the impurity induced pinning field. We note that in this high $T$ regime, we expect no difference between
the uniform and local quadrupolar susceptibilities. These induced quadrupoles around the impurity site 
preserve the local point group symmetry, but lead to
symmetry allowed renormalization of the effective gyromagnetic ratio for $^{23}$Na, causing
a temperature dependent Knight shift. The inverse Knight shift thus scales as $\chi^{-1}_Q(T)$, which
has a linear $T$ dependence in the Curie-Weiss regime.

\noindent ii) At intermediate temperatures ($T=3 T_c, 2 T_c, T_c$ in Fig.~\ref{fig:nmr}{\bf a}),  
we find that in addition to the temperature dependent Knight shift 
of the dominant
peak, the NMR spectrum exhibits broadening as we cool, and further develops an asymmetry. This asymmetry may be
seen via the growing skewness shown in Fig. \ref{fig:nmr}{\bf c} (see Methods). These effects
arise due to the development of a growing inhomogeneous distribution of static quadrupolar droplets in the presence of the 
pinning due to impurity Na$^+$ sites. 
As seen from Fig.~\ref{fig:SchematicDistortion}, an isolated Na$^+$ impurity only leads to an
$A_{1g}$ deformation of the surrounding oxygen octahedron; as discussed above,
this does not lead to NMR peak splitting but only to a Knight shift.
Instead, the broadening and asymmetry in our theory arises from regions harboring pairs (or more)
of neighboring Na impurities which lowers the point group symmetry at the Na site and leads to peak splittings.

\noindent iii) Finally, as we cool below $T_c$ ($T=0.75 T_c, 0.5 T_c, 0.1 T_c$ in Fig.~\ref{fig:nmr}{\bf a}, 
the Knight shift as defined through the first moment changes its slope, having a weaker
temperature dependence below $T_c$.
We can understand this as the growth of $\chi_Q(T)$ getting cut off below $T_c$ due to the onset of octupolar order
as seen from Fig.~\ref{fig:Chi}, while the weaker impact of the octupolar order leads to a smaller slope. 
In addition, as we go below $T_c$,
the NMR spectrum becomes extremely broad. In our calculations,
this broadening reflects the presence of parasitic dipoles induced by the octupolar symmetry breaking.
The impact of octupolar ordering is also reflected in the asymmetry of the spectrum. As shown in Fig ~\ref{fig:nmr}{\bf c}, 
the skew of the spectrum, defined as the third moment of the distribution, shows a singular increase below $T_c$, 
so that the octupolar ordering below $T_c$ leads to a 
more asymmetric spectrum. This is in good agreement with recent experiments \cite{vesna2022}.
We thus conclude that these NMR experiments are indirectly probing the onset of ferro-octupolar symmetry breaking.

\section*{Discussion}





In summary, our work introduces the key concept of Janus impurities to refer to local impurity probes of 
non-Kramers doublets and their ordering.
We have carried out a comprehensive study of how such Janus impurities break the non-Kramers degeneracy
by inducing local strains, and how they can nevertheless serve as useful probes of multipolar orders.

As an example, we have studied the osmate double perovskites which feature competing octupolar and quadrupolar moments.
Our work goes beyond previous studies in several respects, and makes direct contact
with a body of recent experiments.
(1) We have presented DFT computed phonon spectra in the osmates which are 
in good agreement with 
neutron scattering results, making the case that these osmates are intrinsically stable cubic materials.
(2) We showed DFT results for how a doped impurity modifies the local structure and lowers the point group symmetry, 
leading to not only $E_g$ strain but also $T_{2g}$ strains, and 
presented the associated Landau theory for its impact on the local non-Kramers moment. 
(3) The above progress led us to a simplified one-parameter model for doped impurities. The resulting impact on
Tc suppression, computed using Monte Carlo simulations, was shown to explain the data from
recent $\mu$SR experiments.
(4) Our Landau theory also showed that the $T_{2g}$ strains, found in DFT, can nucleate local 
dipole moments in the ground state
even though the cubic system has no ground state dipole moments (only quadrupolar and octupolar).
(5) Finally, we presented a symmetry based approach to modelling the NMR spectra of doped $^{23}$Na,
keeping track of the full inhomogeneous distribution
of the local Os moments. This allows us to make direct contact with the temperature
evolution of the NMR 
Knight shift and spectral features seen in the osmates both above and below Tc.
In a broader sense, our work sets the framework for understanding NMR in an inhomogeneous multipolar environment in 
diverse quantum materials.

By combining {\it ab initio} methods, Landau theory, MC simulations, and
modelling of NMR spectra, we have thus shown how local impurity strains can suppress the 
ferro-octupolar $T_c$, and how our work explains
recent $^{23}$Na NMR results on {\bnco} for doping $\delta \! \ll\! 1$. Our results
lends further support to the scenario of octupolar order below $T_c$ in these $d$-orbital materials,
and show that Janus impurities can reveal useful information about multipolar orders.

We note that the Na$^{+}$
impurity (i.e., substitution of Ca$^{2+}$ by Na$^{+}$) 
not only produces strong local strain fields, but would
also lead to local charge doping: each impurity could
potentially dope $1/6$-hole onto each of the six neighboring Os. Our proposal here is that the local
strain can play a highly significant role in accounting for the drop in $T_c$; the detailed exploration 
of charge doping effects and longer-range strain fields is a subject for future study.
Our {\it ab initio} computations find that similar local
strain fields are generated in {\bsco} due to isovalent substitution of Ca$^{2+}$ by
Sr$^{2+}$.
Experimental studies of {\bsco} and {\bcmo}, perhaps using $^{87}$Sr NMR or $^{25}$Mg NMR, would thus provide a route to 
distinguish the effects of strain from charge doping. Sr or Mg doping may also allow one to study the
QCP where the octupolar $T_c$ is suppressed to zero. The impact of charge doping with Na$^+$ substitution
going from $\delta=0$ to $\delta=1$ \cite{vesna2022} deserves a separate careful investigation.
In addition,
NMR studies on the undoped $d^2$ osmate double perovskites would also be extremely valuable to 
explore spectral signatures in the absence of strain fields. Finally, while we have focussed here on $d$-orbital multipoles, most of our
results can also be directly translated to $f$-orbital
candidates with ordered multipole moments such as PrV$_2$Al$_{20}$ 
\cite{Sakai_JPSJ2011,Sato_PRB2012,Nakatsuji_PRL2014,Hattori2016,Freyer2018,SBLee2018,Patri2019}, 
or to systems with fluctuating multipoles such as spin-ice materials \cite{spinice_Changlani_npjQM2022}.

\section*{Methods}

{\bf Ab Initio Calculations:} 
The {\it ab initio} calculations are carried out in the plane wave pseudo-potential basis, as implemented in Vienna 
ab-initio Simulation Package \cite{vasp1,vasp2}, choosing the generalized gradient approximation (GGA) 
to describe the exchange-correlation functional.
We have also examined effect of correlation beyond the GGA level by doing calculations
within Hubbard $U$ supplemented GGA+$U$ calculations, with $U$ value of $1$-$2$\,eV
applied at Os site; we have found that the results are qualitatively unchanged.
The effect of SOC which is especially important at the 5d Os sites is taken into account
through GGA+$U$+SOC calculations.
The cutoff energy of the plane-wave basis is chosen 
to be $600$\,eV which is found to be sufficient to achieve convergence in self-consistent field (SCF) calculations. 
Converged k-mesh of $8\times 8\times 8$ is used for SCF calculation with tight energy convergence threshold of 
$10^{-8}$ eV. Relaxations of the crystal structures are carried out with respect to internal atomic coordinates
as well as cell volume using a convergence threshold of $10^{-5}$ eV for total energy and
$10^{-3}$ eV/{\AA}
for maximum force/atom. Phonon properties are studied within the formulation of density functional perturbation theory (DFPT)
as implemented in VASP. The phonon band structures were obtained by Fourier interpolation of the real-space force
constants using PHONOPY \cite{phonopy} code, and plotted along the high-symmetry momentum points of the BZ.

{\bf Monte Carlo Simulations:}
Monte Carlo (MC) simulations were carried out on a $14\times14\times14$ fcc lattice (2744 spins)
with periodic boundary conditions. The spins on the fcc sites form one sublattice of a cubic lattice
which represent the Os pseudospins. Impurity sites are randomly selected from the second sublattice
of the cubic lattice, representing substitution of Ca by Na or Sr, and the effect of this impurity
in the simulations was mimicked by applying strain fields on the six neighboring
Os atoms (in accordance with Eqs. \eqref{strainfieldeq}-\eqref{streqz}).
For Os sites which happen to be adjacent to multiple impurities, the corresponding strain fields are (vectorially) added. 
The MC simulations were thermalized over $10^6$ MC sweeps, and measured over $5\times 10^5$ sweeps. 
Results for each impurity concentration are averaged over $20$ disorder realizations for $T_c$. For
computing the NMR spectra for $\delta=0.1$, we average over $100$ disorder configurations to capture the
spectra over many Na nuclei.

{\bf NMR Spectrum Calculation: }
To obtain the spectrum shown in Fig. \ref{fig:nmr}{\bf a}, we solve for the eigenvalues of Eq. \eqref{localnmr} (a $4\times4$ matrix), and associate the differences between successive eigenvalues as the observed frequencies in an NMR measurement. We thus have three frequencies associated with a given impurity atom. We collect the frequencies from all the impurity atoms in 100 disorder configurations 
(total $27440$ impurity spectra for $\delta\!=\!0.1$), 
averaged over 300 Zeeman field directions to produce a data set that mimics that obtained through a powder sample NMR measurement. To obtain a simulated spectrum, the binned data is then smoothed using a Kernel Density Estimate (KDE) with a bandwidth of 
$2 \times 10^{-4} \omega_0$. The mean is defined as the first moment of the distribution, $\la \omega \ra$, and the skew is defined as the third moment, $\la (\omega - \la\omega\ra)^3 \ra$. 
The histogram from the raw data that was used to produce Fig. \ref{fig:nmr} is given in Supplementary Note 6.

\section*{Data availability}

The authors declare that the data supporting the findings of this study are available within the paper and its supplementary information files.

\section*{Code availability}
{The Monte Carlo codes used in this study are available from
\href{https://github.com/sreekar-voleti/SpinMC_more.jl}{https://github.com/sreekar-voleti/SpinMC{\_}more.jl}}

\section*{Competing interests}
{The authors declare no competing financial or non-financial interests.}

\section*{acknowledgments}
We thank Vesna Mitrovic for useful discussions about their NMR results.
This research was funded by a Discovery Grant from the Natural Sciences and Engineering Research Council 
of Canada RGPIN-2021-03214 (SV, AP), DST India (SB and TS-D), and
a SERB-India Vajra Fellowship VJR/2019/000076 (AP, TS-D, SB). 
TS-D acknowledges a J.C.Bose National Fellowship (grant no. JCB/2020/000004) 
for funding. Monte Carlo and NMR spectra 
computations were carried out on the Niagara supercomputer at the SciNet 
HPC Consortium and the Digital Research Alliance of Canada.

\section*{Contributions}
AP, TS-D, and SV designed the research. SV and KP performed the numerical computations. All authors 
(SV, KP, AP, SB, and TS-D) were involved in discussing
the data and results. SV, AP, and TS-D contributed to the writing of the manuscript with inputs from all authors.

\bibliography{octupolar}

\begin{table}[b]
\centering
\begin{tabular}{|P{2.5cm}|P{1.2cm}|P{1.2cm}|P{1.2cm}|P{1.2cm}|}
 \hline
Exchange (meV) & $K_{\rm o}$ & $K_1$ & $K_2$ & $T_c$(K) \\ [0.5ex] 
 \hline
Model A & -1.0  & 0.1 & 0.05 & 40 \\ 
 \hline
Model B & -1.0  & 0.4 & 0.2 & 40 \\ 
 \hline
Model C & -2.98  & 1.48 & -0.61 & 115\\ 
 \hline
Model D & -1.91  & 0.16 & 0.42  & 78\\ 
 \hline
\end{tabular}
\caption{{\bf Exchange model parameters.} Models for the multipolar exchange couplings on the ideal fcc lattice, with a
ferro-octupolar exchange $K_{\rm o}$ and quadrupolar exchanges $K_1, K_2$. Models A and
B are from microscopic tight-binding models studied in ref.~\cite{Voleti2021}. The exchange couplings for
Model C \cite{pourovskii2021} and Model D (this work, see Supplementary Note 4) are derived 
using two different DFT-based methods; the $T_c$ for these two models is much larger than experimental observations
which may reflect a combination of stronger quantum fluctuations (due to larger quadrupolar
exchanges) and possible limitations of DFT in accurately extracting small 
$\sim\! 1$\,meV exchange scales.}
\label{table:1}
\end{table}

\end{document}